\documentclass[aps,preprint,floatfix,nofootinbib,showpacs]{revtex4}
\usepackage{graphicx}
\usepackage{color}
\newcommand{\U}{{\cal U}}
\begin{document}

\title{Global Study of Electron-Quark Unparticle Interactions}
\renewcommand{\thefootnote}{\fnsymbol{footnote}}
\author{ 
Chun-Fu Chang$^{1}$,
Kingman Cheung$^{2,1,3}$, 
and Tzu-Chiang Yuan$^{4}$
}
\affiliation{$^1$Department of Physics, National Tsing Hua University, 
Hsinchu 300, Taiwan
\\
$^2$Division of Quantum Phases \& Devices, School of Physics, 
Konkuk University, Seoul 143-701, Korea \\
$^2$Department of Physics, National Tsing Hua University, 
Hsinchu 300, Taiwan
\\
$^3$Physics Division, National Center for Theoretical Sciences,
Hsinchu 300, Taiwan
\\
$^4$Institute of Physics, Academia Sinica, 
Nankang, Taipei 11529, Taiwan
}

\renewcommand{\thefootnote}{\arabic{footnote}}
\date{\today}

\begin{abstract}
  We perform a global fit on parity-conserving electron-quark
  interactions via spin-1 unparticle exchange.  Besides the peculiar
  features of unparticle exchange due to non-integral values for the
  scaling dimension $d_\U$ and a non-trivial phase factor $\exp
  (-id_\U \pi)$ associated with a time-like unparticle propagator, the
  energy dependence of the unparticle contributions in the scattering
  amplitudes are also taken into account.  The high energy data sets
  taken into consideration in our analysis are from (1) deep inelastic
  scattering at high $Q^2$ from ZEUS and H1, (2) Drell-Yan production
  at Run II of CDF and D\O, and (3) $e^+e^-\to\rm hadrons$ at LEPII.
  The hadronic data at LEPII by itself indicated a $3-4$ sigma
  preference of new physics over the Standard Model. However, when all
  data sets are combined, no preference for unparticle effects can be
  given.  We thus deduce an improved 95\% confidence level limit on
  the unparticle energy scale $\Lambda_\U$.
\end{abstract}


\maketitle

\section{Introduction}

Scale invariance is a very powerful symmetry in both physics and
mathematics even though it is not an exact symmetry in the quantum
world.  As one considers the renormalizable theories of interacting
elementary particles, the classical scale invariance is broken either
explicitly by some dimensional mass parameters in the theory or
implicitly by renormalization effects.  For example, the scale
invariance is broken in the Lagrangian of the Standard Model (SM) at
tree level just by a negative mass squared term in the Higgs
potential.  Also in massless quantum chromodynamics (QCD),
renormalization effects can give rise to scaling violations phenomena
through the effects of running couplings and masses. Even though one
has only an approximate scale invariance at low energy physics that is
described well by the SM and QCD, this cannot prevent one from
imagining there might be an exact scale invariant sector at a higher
energy scale that has not yet been probed by experiments.

Georgi \cite{unparticle}, motivated by the Banks-Zaks theory \cite{Banks-Zaks},
suggested that a scale invariant hidden sector with a nontrivial 
infrared fixed point could exist 
at high enough energy. 
Such a scale invariant hidden sector may couple strongly or weakly within
 itself, but its interactions with the SM fields are presumably weak such that
 an effective field theory approach can be employed.
Though we are ignorant of this hidden sector above this high energy scale, 
we still can use the approach of effective field theory to probe its 
low energy effects at the TeV scale.
 
In Georgi's scheme \cite{unparticle}, the scale invariant sector, or denoted as
Banks-Zaks ($\cal BZ$) sector, can interact with the SM fields through
a messenger sector at a high mass scale $M_\U$.  Below this high mass
scale $M_\U$, the non-renormalizable operators are suppressed by
inverse power of $M_\U$ and schematically represented in the following 
form \cite{unparticle}
\begin{equation}
   \frac{ 1 } {M_\U^{d_{\mathrm SM} + d_{\cal BZ}-4}} \, 
   {\cal O}_{\mathrm SM} \, {\cal O}_{\cal BZ}  \; ,
\label{genericop}
\end{equation}
where ${\cal O}_{\mathrm SM}$ and ${\cal O_{BZ}}$ represent the SM and
$\cal BZ$ fields with scaling dimensions $d_{\mathrm SM}$ and $d_{\cal
  BZ}$ respectively.  As one scales down the theory from the higher
scale $M_\U$, this hidden sector may flow to an infrared fixed point
at the scale $\Lambda_\U$.  Georgi coined this hidden stuff as
`unparticle' $\U$.  Below $\Lambda_\U$ one needs to replace the
operators in (\ref{genericop}) with a new set of operators having the
form
\begin{equation}
    C_{\cal O_U} \frac{ 1 } {M^{d_{\mathrm SM} + d_{\cal BZ}-4}_\U 
\Lambda_\U^{ d_\U - d_{\cal BZ}} } \,
   {\cal O}_{\mathrm SM}\, {\cal O}_\U \;, \smallskip
\label{effectiveop}
\end{equation} 
where ${\cal O_U}$ is the unparticle operator with the scaling dimension $d_\U$ 
and $C_{\cal O_\U}$ is the coefficient function. 
For an interacting scale invariant theory, the scaling dimension $d_\U$,
unlike the case for a canonical boson or fermion field, 
is not necessarily an integer or half-integer.
Besides, the unparticle operator ${\cal O_U}$ 
with a general non-integral scaling dimension $d_\U$ has a mass spectrum 
looked like a $d_\U$ number of invisible massless particles \cite{unparticle}.
Explicit list of SM invariant operators of the form Eq.(\ref{effectiveop}) 
was written down in Ref.~\cite{Chen-He}.

There have been many phenomenological studies relevant to unparticle
physics in the past several years.  A recent summary on the phenomenology 
and a more complete 
list of references can be found in Ref.~\cite{tc}.  
However, there has not been a
precise study of global constraint on unparticle interactions,
except for an approximated estimate of $\Lambda_\U$
\cite{CKY-long,bander},
based on a global study of 4-fermions contact interactions
\cite{Cheung}.  Such a naive estimate could not take into
account the energy dependence of the effective 4-fermions interactions due to 
virtual unparticle exchange in the event-by-event basis.
  In this work, we redo the analysis from scratch,
in which full energy dependence in the event-by-event basis
is taken into account
in each experimental data set. In other words, our results are more
accurate and valid.
We analyze the
parity-conserving $\ell \ell q q$ spin-1 unparticle interactions
against the high energy data sets on neutral current $\ell\ell q q $
interactions, including deep inelastic scattering at high $Q^2$ from
ZEUS and H1, Drell-Yan production at Run II of CDF and D\O, and
$e^+e^-\to\rm hadrons$ at LEPII.  In the analysis, we found that the
hadronic data at LEPII showed a $3-4$ sigma preference of new physics
over the SM. However, when all data sets are combined, the data showed
no preference.  Thus, we obtain 95 \% C.L. limits on the unparticle
energy scale $\Lambda_\U$.

One can also consider spin-0 unparticle exchanges via operators such as
\[
 \frac{\lambda m_f}{\Lambda_\U^{d_\U } } O_\U \bar f f , \qquad
 \frac{\lambda}{\Lambda_\U^{d_\U} } (\partial^\mu O_\U) \bar f \gamma_\mu f 
\;.
\]
The former scalar unparticle exchange is expected to be proportional
to the Yukawa coupling of the fermion.  In the high energy data sets that
we study such exchanges are negligible. On the other hand, the latter
scalar unparticle exchange is non-negligible because of the
derivative coupling.  The limits on such a scalar unparticle exchange
were obtained in Ref. \cite{bander}. 
The limits on scalar unparticle from Bander {\it et al}. \cite{bander}
range from $0.46$ TeV for $d_\U = 1.9$ to $2.1$ TeV for $d_\U = 1.1$.
Nevertheless, the limits are
in general an order of magnitude smaller than those obtained
for vector unparticle exchange for the same $d_\U$.  Limits for spin-1 
vector exchanges are also in general more stringent than those of spin-2
exchanges.

The organization of the paper is as follows. In Sec. II we present
the formulation of unparticle physics and the parameterization of the 
electron-quark unparticle interactions that are needed in our study.  In
Sec. III we describe the data sets that are used in this global analysis,
including the deep inelastic scattering at high $Q^2$ from ZEUS and H1,
the Drell-Yan production at Run II of CDF and D{\O}  as well as the 
total hadronic
cross section $\sigma_{\rm had}$ at LEPII. We present
our analysis and results in Sec. IV and conclude in Sec. V.

\section{Formulation of Unparticle Interactions}

One way to probe the existence of unparticle physics is via the
interference effects between the pure SM amplitudes 
and the similar ones with unparticle exchange \cite{georgi2,CKY-short}.
It is interesting to note that 
the unparticle propagator has a peculiar phase factor
$\exp(-i d_\U \pi)$ associated with the time-like momentum
transfer \cite{georgi2,CKY-short}. 
This complex phase in the unparticle propagator will give rise to
non-trivial interference with
the propagators of SM particles without this peculiar phase factor. 
In this study, we investigate $\ell\ell qq$
interactions by exchange with SM gauge bosons and vector unparticles.
   
The effective interactions of the vector unparticle operator with 
SM fermions is parameterized by
\begin{equation}
{\cal L}_{\mathrm{eff}}  \ni \lambda_1 \frac{1}{\Lambda_\U^{d_\U - 1} }\,
 \bar f \gamma_\mu f \,
O_\U^\mu \;, 
\end{equation}
where $f$ denotes a SM fermion field and $\lambda_1$ is the 
dimensionless effective coupling.
The propagator of vector unparticle is \cite{georgi2,CKY-short,CKY-long}
\begin{equation}
\left[ \Delta_F (P^2) \right]_{\mu\nu} =
         Z_{d_\U} \, (-P^2)^{d_\U -2} \, 
         \pi_{\mu\nu}(P) \;, 
\end{equation}
with
\begin{eqnarray}
Z_{d_\U}  &=& \frac{A_{d_\U}} { 2 \sin (d_\U \pi) } \qquad ~\; , \\
\pi^{\mu\nu}(P) &=& - g^{\mu \nu} + \frac{P^\mu P^\nu }{ P^2} \quad , \label{pi}
\end{eqnarray} 
and $A_{d_\U}$ is given by
\[
A_{d_\U}={16\pi^{5 \over 2} \over (2\pi)^{2{d_\U}}}
       { \Gamma({d_\U}+{1\over
       2})\over\Gamma({d_\U}-1)\Gamma(2\,{d_\U})}  \; .
\] 
We need the following different treatment about the factor of $(- P^2)^{d_\U - 2}$ in the propagator
\begin{equation}
\label{branchcut}
 (-P^2)^{d_\U -2}=\left \{
\begin{array}{lcl}
|P^2|^{d_\U -2}   & \quad & \hbox{if $P^2$ is negative and real, } \\
|P^2|^{d_\U -2} e^{-i d_\U \pi} & & \hbox{for positive $P^2$ with 
an infinitesimal $i0^+$} . 
\end{array} \right.
\end{equation}
Therefore, the $s$-channel propagator has the nontrivial phase factor 
but the $t$- and $u$- channel propagators do not.
We have imposed the requirement that the spin-1 unparticle propagator
is transverse, as indicated by the tensor structure of the projection operator 
in Eq. (\ref{pi}).  
If the more restricted conformal 
invariance is assumed for the unparticle sector, the following 
slightly more complicated tensor structure for the projection 
operator must be used  \cite{GIR}:
\begin{equation}
\pi^{\mu\nu}(P) = - g^{\mu \nu} + \frac{2 \left( d_\U - 2 \right)}{\left( d_\U - 1 \right)} 
\frac{P^\mu P^\nu }{ P^2} \quad . 
\label{pi2}
\end{equation}
However, in the case of massless external fermions, both forms of the 
projection operators lead to the same effective $\ell\ell qq$ operators. 
A similar remark
can be made to the spin-2 unparticle propagator as well.
The second term in (\ref{pi2}) does give rise to nontrivial effects in the case
where the external fermion mass cannot be ignored, for example in the
$B_s \overline B_s$ system \cite{lee}. 
As discussed by Georgi in his original works \cite{unparticle,georgi2}, 
the key feature of unparticle is scale invariance.
However, no one has been able to find a physically sensible
interacting theory which has 
scale invariance but not conformal.  
\footnote{
We note that a non-conformal but scale-invariant 
two-dimensional elasticity theory was constructed
in Ref. \cite{cardy}. However the theory is non-unitary.
}
Therefore, one generally expects there is a close relation 
between scale
invariance and being conformal.
If the scale invariance is extended to be conformal,
the value of $d_\U$ imposed by unitarity has to be larger than 3 for a 
vector unparticle 
propagator \cite{mack}. 
Hence, for completeness, in our numerical analysis presented in Sec. IV, 
we will consider the range of $d_\U$ varying from 1 to 4.

When the scaling dimension $d_\U$ is larger than 2, the propagator
factor was shown to be modified \cite{sannino}.
The form of the propagator depends on the UV completion, because
counter terms have to be included such that the dependence on
the UV scale can become mild.  In \cite{sannino}, it was 
shown that the dependence on UV scale is only logarithmic when
$d_\U \to 2$.  
We anticipate that when appropriate counter terms are added the
dependence on the UV scale could be more severe than being logarithmic
for other values of $d_\U > 2$, namely that the divergence is linear
for $d_\U =3$ and more than linear for $d_\U > 3$.

To be specific, consider the parton subprocess of $q\bar{q} \to e^+e^-$. 
After including 
the virtual exchange of spin-1 unparticle, the amplitude squared 
(without averaging initial spins or colors) is given by
\begin{equation}
\sum |{\cal M}|^2 = 4 \hat{u}^2 \left( |M^{eq}_{LL}(\hat{s})|^2 + 
     |M^{eq}_{RR}(\hat{s})|^2 \right )
  + 4 \hat{t}^2 \left( |M^{eq}_{LR}(\hat{s})|^2 + 
|M^{eq}_{RL}(\hat{s})|^2 \right ) \;,
\end{equation}
with
\begin{equation}
 M^{eq}_{\alpha\beta}(\hat s)  =  
  \     \lambda_1^2 Z_{d_\U}  \frac{1}{\Lambda_\U^2} 
 \left (- \frac{\hat s}{\Lambda_\U^2} \right)^{d_\U-2} 
+ \frac{ e^2 Q_e Q_q}{ \hat s}
  + \frac{e^2 g^e_\alpha g^q_\beta}{ \sin^2 \theta_{\rm w}
  \cos^2\theta_{\rm w}
 }\, \frac{1}{\hat s - M_Z^2+iM_Z\Gamma_Z} \; ,
\label{reduced}
\end{equation}
where the $M^{eq}_{\alpha\beta}(\hat s)$ is the tree level reduced amplitude
and the subscripts stand for the chiralities of electrons($\alpha$) and
quarks($\beta$). 
In the reduced amplitude, 
$\hat{s}$ is the subprocess center-of-mass energy squared, 
$g_L^f = T_{3f} - Q_f \sin^2 \theta_{\rm w}$ and
$g_R^f = - Q_f \sin^2 \theta_{\rm w}$ with $T_{3f}$ and $Q_f$ being 
the third component of the $SU(2)_L$ and the electric charge of the
fermion $f$ in units of the proton charge respectively, 
$\sin \theta_{\rm w}$ is the Weinberg angle 
and $e^2=4\pi \alpha_{\rm em}$.

To facilitate our analysis in the next section, we introduce a parameter 
$\epsilon$ by letting 
\begin{equation}
\epsilon \equiv \left( \frac{\lambda_1}{\Lambda^{d_\U-1}_{\U}} \right)^2 \;,
\label{eps}
\end{equation}
in order to get a term similar to four-fermion contact interactions.
By setting $\epsilon = 0$ we recover the original SM amplitude.
For simplicity, we assume the unparticle sector is flavor- and chirality-blind.
For the $s$-channel amplitude $\hat s > 0$, we have to insert the
phase factor $\exp(-i d_\U \pi)$ 
into the unparticle four-fermion contact term and it will interfere with 
the other 
terms from the photon and $Z$ boson exchange in a peculiar way for 
non-integral $d_\U$. 
On the other hand, 
when we consider non-$s$-channel processes, like $eq \to eq$ at HERA, 
we must replace the Mandelstam variable $\hat{s}$ with $\hat{t}$ or $\hat{u}$ 
and drop the Breit-Wigner width $iM_Z\Gamma_Z$ 
in the $Z$ boson propagator.
Equation~(\ref{reduced}) augments the SM amplitudes with the unparticle physics 
contribution in the form of an extra four-fermion contact term. We also 
note that for non-integral
$d_\U$ the contact term in Eq.(\ref{reduced}) has a nontrivial energy 
dependence through the factor
$\left (- \frac{\hat s}{\Lambda_\U^2} \right)^{d_\U-2}$. 
Such a contact term is widely used in the model construction associated 
with new physics.
Through a global fit of data sets we can judge if new physics can be 
discerned in current experiments through the contact term \cite{Cheung}.
   
\section{Experimental Data Sets}

\subsection{HERA data}

We adopt the neutral current deep inelastic scattering (DIS) $e^- p$
data from ZEUS and $e^+ p$ scattering data from H1 \cite{hera}, both
measured at a center-of-mass energy $\sqrt{s} \approx 318$ GeV with an
integrated luminosity of 169 pb$^{-1}$ and 65.2 pb$^{-1}$,
respectively.
 
The commonly used kinematic variables in DIS are $x$, $y$, and $Q^2$, which
are related by
\begin{equation}
x = \frac{Q^2}{  2p\cdot (k-k') } \; , \qquad y = \frac{Q^2 }{ sx } \,.
\end{equation}
Here $k$ and $k'$ are the four-momentum of the incoming and outgoing leptons, 
$p$ is the four-momentum of the proton and
$Q^2$ is minus the square of the momentum transfer
\begin{equation}
Q^2 = -(k-k')^2 = sxy \,.
\end{equation}

On analyzing ZEUS data, we use the reduced cross section $\tilde{\sigma}$ 
defined as 
\begin{equation}
\tilde{\sigma}(e^- p) = \frac{xQ^4}{2\pi \alpha_{\rm em}^2} \frac{1}{1 + (1-y^2)} \frac{d^2\sigma(e^- p)}{dxdQ^2}\, ,
\end{equation}
with the double differential cross section 
\begin{eqnarray}
\frac{d^2\sigma(e^- p)}{dxdQ^2} &=&
 \frac{1}{16\pi}\; \Biggr \{
\sum_q f_q(x) \,\biggr [
    |M^{eq}_{LL}(\hat{t})|^2 + |M^{eq}_{RR}(\hat{t})|^2  +
    (1-y)^2 (|M^{eq}_{LR}(\hat{t})|^2 + |M^{eq}_{RL}(\hat{t})|^2) \biggr ]
\nonumber  \\
&+&
\sum_{\bar q} f_{\bar q} (x) \, \biggr [
     |M^{eq}_{LR}(\hat{t})|^2 + |M^{eq}_{RL}(\hat{t})|^2  + 
    (1-y)^2 (|M^{eq}_{LL}(\hat{t})|^2 + |M^{eq}_{RR}(\hat{t})|^2) \biggr ]
    \, \Biggr \} \; ,  \label{nccc}
\end{eqnarray}
where
$f_{q/\bar q}(x)$ are parton distribution functions. We use CTEQ (v.6)
parton distribution functions wherever they are needed. The reduced
amplitudes $M^{eq}_{\alpha\beta}$ are given by Eq. (\ref{reduced}).
On the other hand, on analyzing the H1 data we use the single differential
cross section $d\sigma(e^+ p)/dQ^2$ by interchanging $( LL
\leftrightarrow LR, RR \leftrightarrow RL)$ in the reduced amplitudes 
$M^{eq}_{\alpha\beta}$ in Eq. (\ref{nccc}) and then integrating over
the $x$ variable.

We first calculate the reduced cross section $\tilde \sigma$ of the SM
contributions and normalize it to the whole data sets to determine the
overall scale factor $C$, which is pretty close to 1. We then include
the unparticle contributions into $\tilde \sigma$ and multiply it by
the scale factor $C$ determined from the previous step.  For ZEUS
data, the reduced cross section used in the minimization procedure is
given by
\begin{equation}
\label{sign}
\tilde{\sigma}^{\rm th} = C \left( \tilde{\sigma}^{\rm SM} + \tilde{\sigma}^{\rm interf} +
\tilde{\sigma}^{\rm unpart} \right)  \; ,
\end{equation}
where $\tilde{\sigma}^{\rm interf}$ is the interference cross section
between the SM and the unparticle four-fermion interactions and
$\tilde{\sigma}^{\rm unpart}$ is the cross section due to the
unparticle interactions alone.  We then compare the corrected
theoretical values $\tilde{\sigma}^{\rm th}$ with the experimental
results. Similarly, when treating the H1 data, we follow the same
minimization procedure for the reduced cross section with the single
differential one.

\subsection{Tevatron: Drell-Yan Process}

We use the preliminary Run II data of Drell-Yan (DY) production
extracted from the CDF and D\O\ figures \cite{tevatron}.  Both are
measured in the form of $d\sigma/dM_{ee}$, where $M_{ee}$ is the
invariant mass of the electron-positron pair. The double differential
cross section which includes the contributions of the spin-1
unparticle interactions is given by
\begin{equation}
\frac{d^2\sigma}{dM_{ee} dy} = K \frac{M_{ee}^3}{72\pi s} \;
\sum_q f_q(x_1) f_{\bar q}(x_2) \left(
|M^{eq}_{LL}(\hat s)|^2 + |M^{eq}_{LR}(\hat s)|^2 + |M^{eq}_{RL}(\hat s)|^2 + 
|M^{eq}_{RR}(\hat s)|^2  \right ) \; ,
\label{drell-yan}
\end{equation}
where the amplitudes $M_{\alpha\beta}^{eq}$ are given by
Eq. (\ref{reduced}).  In Eq. (\ref{drell-yan}), $\sqrt{s} = 1.96~ \rm
TeV$ is the center-of-mass energy of the $p\bar{p}$ collisions,
$\hat{s} = M_{ee}^2$, $y$ is the rapidity of the electron-positron
pair and $x_{1,2} = M_{ee}\,e^{\pm y}/\sqrt{s}$.  We will numerically
integrate over the rapidity $y$ distribution in our analysis.  The QCD
$K$-factor of Drell-Yan production is known to 1-loop as
\begin{equation}
K = 1 + \frac{\alpha_s(M^2_{ee})}{2\pi} \,\frac{4}{3} \left(
1+ \frac{4}{3}  \pi^2 \right ) \;.
\end{equation}
With this $K$ factor, the overall cross section normalization
agrees with the Tevatron data in the vicinity of the $Z$-peak.  

\subsection{LEP data}

The LEP Electroweak Working Group (LEPEW) combined the data of total
hadronic cross section from the four LEP collaborations at energies
from $130$ GeV to $209$ GeV \cite{lep}. 
In the LEPEW report, they noted
that the ratio of the measured cross sections to the SM expectations,
averaged over all available energies, showed an approximate $1.7 \sigma$
excess. We also see this effect in our fits.

In the report, both the experimental cross sections and the SM
predictions are given.  Since the predictions given in the report do
not take into account unparticle interactions, we do the calculation
by first normalizing our tree-level SM results to the predictions
given in the report and then multiplying this scaling ratio 
to the new cross sections that include the SM and the unparticle interactions.

At leading order in the electroweak interactions, the total hadronic cross
section for $e^+e^-\to q\bar q$, summed over all flavors $q=u,d,s,c,b$, is
given by
\begin{equation}
\sigma_{\rm had} \left( s \right) = K\, \sum_q \frac{s}{16\pi} \biggr[
|M_{LL}^{eq}(s)|^2 + |M_{LR}^{eq}(s)|^2 + |M_{RL}^{eq}(s)|^2 +
|M_{RR}^{eq}(s)|^2 \biggr] \; ,
\label{sig_had}
\end{equation}
where $M_{\alpha\beta}^{eq}$ is given by Eq. (\ref{reduced}).
The prefactor $K$ is the QCD correction known to 3-loop as
\begin{equation}
K=1+\left( \frac{\alpha_s}{\pi} \right) + 1.409\left( \frac{\alpha_s}{\pi} \right)^2 
- 12.77 \left( \frac{\alpha_s}{\pi} \right)^3 \; .
\end{equation}

\section{Analysis and Results}

Because of severe experimental constraints on intergenerational transitions like
$K\to\mu e$
we restrict our discussions to first generation contact terms. Only
where required by particular data (e.g. the muon sample of Drell-Yan
production at the Tevatron) shall we assume universality of contact terms 
between $e$ and $\mu$.

The effect of unparticle in the scattering amplitude for $q \bar q \to
e^+ e^-$ is explicitly given in Eq. (\ref{reduced}) and similar
formulas for other cases. In order to linearize the fitting procedures
we use $\epsilon = \lambda_1^2 / \Lambda_\U^{2 d_\U -2}$ of
Eq. (\ref{eps}).  The deviation from the SM is parameterized by
$\epsilon Z_{d_\U} ( - \hat s)^{d_\U -2}$.  The predictions by the
model are then compared with the experimental data from Tevatron, LEP
and HERA, as described above. We then calculate the $\chi^2$ as a
function of the parameter $\epsilon$.  We use MINUIT to minimize the
$\chi^2$ function with respect to $\epsilon$, so as to obtain the
minimum $\chi^2_{\rm min}$, which occurs at a particular
$\epsilon_{\rm min}$.
At the same 
time, we can calculate the 95\% range of $\epsilon$ given by the following
\begin{equation}
 0.95 = \frac{ \int_0^{\epsilon_{95}} \; \exp \left[ - \Delta \chi^2 (\epsilon) \right] \;
  d \epsilon }
{ \int_0^{\infty} \; \exp \left[ - \Delta \chi^2 (\epsilon) \right] \;
  d \epsilon } \;,
  \label{eps95}
\end{equation}
where $\Delta \chi^2(\epsilon) \equiv \chi^2(\epsilon) - \chi^2_{\rm min}$.
Here we use the fact that $\epsilon$ only takes on positive
physical values.

\begin{table}[t!]
\caption{\label{table1}
Fitted values of $\epsilon \equiv 
\lambda_1^2/ \Lambda_\U^{2 d_\U -2 }$ of each experimental set and 
the combined set.  The 95\% C.L. lower limit on $\Lambda_\U$ for each
$d_\U$ is obtained by choosing $\lambda_1 = 1$ in the value 
of $\epsilon_{95}$ defined by Eq. (\ref{eps95}).}
\medskip
\begin{tabular}{c|cccc|c}
\hline
\hline
$d_\U$ & \multicolumn{4}{c|}{Fitted parameter $\epsilon \equiv 
\frac{\lambda_1^2}{\Lambda_\U^{2 d_\U -2 } }$}  &
   $\Lambda_\U$ ($\lambda_1=1$) \\
 &  Tevatron DY & HERA DIS & LEP $q\bar q$ & Combined &  (TeV) \\
\hline
\small
$1.1$ &  $\left(0.79^{+0.61}_{-0.58} \right )\times 10^{-3}$ 
      &  $\left(-0.035^{+0.16}_{-0.15} \right )\times 10^{-3}$ 
      &  $\left(-2.65^{+0.74}_{-0.65} \right )\times 10^{-3}$ 
      &  $\left(-0.023 \pm 0.15 \right )\times 10^{-3}$  
      &   $5.5 \times 10^{14}$ \\
$1.3$ &  $\left(0.52^{+0.39}_{-0.34} \right )\times 10^{-3}$ 
      &  $\left(-0.24^{+1.00}_{-0.98} \right )\times 10^{-4}$ 
      &  $\left(-0.83^{+0.21}_{-0.18} \right )\times 10^{-3}$ 
      &  $\left(-0.18^{+0.98}_{-0.96} \right )\times 10^{-4}$  
      &  $1.7 \times 10^{3}$ \\
$1.5$ &  $\left(0.00 \pm 0.17 \right )\times 10^{-3}$ 
      &  $\left(-0.11^{+0.57}_{-0.56} \right )\times 10^{-4}$ 
      &  $\left(2.81^{+0.46}_{-0.55} \right )\times 10^{-4}$ 
      &  $\left(-0.23^{+0.84}_{-0.76} \right )\times 10^{-4}$  
      &  $7.3$ \\
$1.7$ &  $\left(-0.31^{+0.20}_{-0.30} \right )\times 10^{-4}$ 
      &  $\left(-0.015 \pm 0.25 \right )\times 10^{-4}$ 
      &  $\left(0.40^{+0.089}_{-0.10} \right )\times 10^{-4}$ 
      &  $\left(-0.024^{+0.16}_{-0.72} \right )\times 10^{-4}$  
      &  $2.1$ \\
$1.9$ &  $\left(-0.15^{+0.11}_{-0.12} \right )\times 10^{-5}$ 
      &  $\left(0.077^{+0.55}_{-0.52} \right )\times 10^{-5}$ 
      &  $\left(0.36^{+0.092}_{-0.11} \right )\times 10^{-5}$ 
      &  $\left(-0.003^{+0.091}_{-0.11} \right )\times 10^{-5}$  
      &  $1.7$ \\
\hline
\hline
$2.1$ &  $\left(0.38^{+0.34}_{-0.30} \right )\times 10^{-6}$ 
      &  $\left(-0.10^{+0.28}_{-0.30} \right )\times 10^{-5}$ 
      &  $\left(-1.16^{+0.34}_{-0.30} \right )\times 10^{-6}$ 
      &  $\left(0.01^{+0.29}_{-0.26} \right )\times 10^{-6}$  
      &  $0.62$ \\
$2.5$ &  $\left(0.16^{+0.15}_{-0.47} \right )\times 10^{-6}$ 
      &  $\left(-0.13^{+0.23}_{-0.28} \right )\times 10^{-5}$ 
      &  $\left(0.82^{+0.14}_{-0.18} \right )\times 10^{-6}$ 
      &  $\left(-0.27^{+0.20}_{-0.11} \right )\times 10^{-6}$  
      &  $0.14$ \\
$2.9$ &  $\left(-0.22^{+0.33}_{-1.06} \right )\times 10^{-8}$ 
      &  $\left(-0.10^{+0.18}_{-0.25} \right )\times 10^{-6}$ 
      &  $\left(2.39^{+0.64}_{-0.74} \right )\times 10^{-8}$ 
      &  $\left(-0.05^{+2.40}_{-3.39} \right )\times 10^{-9}$  
      &  $0.16$ \\

\hline
\hline
$3.1$ &  $\left(0.28^{+0.09}_{-0.31} \right )\times 10^{-8}$ 
      &  $\left(0.46^{+1.45}_{-0.91} \right )\times 10^{-7}$ 
      &  $\left(-1.03^{+0.32}_{-0.28} \right )\times 10^{-8}$ 
      &  $\left(-0.06^{+1.50}_{-0.79} \right )\times 10^{-9}$  
      &  $0.10$ \\
$3.5$ &  $\left(0.58^{+0.18}_{-0.28} \right )\times 10^{-9}$ 
      &  $\left(0.34^{+1.77}_{-0.80} \right )\times 10^{-7}$ 
      &  $\left(-1.15^{+0.28}_{-0.22} \right )\times 10^{-8}$ 
      &  $\left(0.58^{+0.18}_{-0.27} \right )\times 10^{-9}$  
      &  $0.066$ \\
$3.9$ &  $\left(0.85^{+0.58}_{-3.30} \right )\times 10^{-11}$ 
      &  $\left(0.24^{+2.27}_{-0.72} \right )\times 10^{-8}$ 
      &  $\left(0.50^{+0.14}_{-0.16} \right )\times 10^{-9}$ 
      &  $\left(0.89^{+0.56}_{-3.22} \right )\times 10^{-11}$  
      &  $0.072$ \\
\hline
\hline
\end{tabular}
\end{table}

\begin{table}[t!]
\caption{\label{table2}
Same as Table \ref{table1} but without the hadronic data set of LEP.}
\medskip
\begin{tabular}{c|c|c}
\hline
\hline
$d_\U$ & Fitted parameter 
    $\epsilon \equiv \frac{\lambda_1^2}{\Lambda_\U^{2 d_\U -2 } }$  &
   $\Lambda_\U$ ($\lambda_1=1$) \\
 &  Tevatron DY + HERA DIS &  (TeV) \\
\hline
\small
$1.1$ &  $\left(0.023 \pm 0.15 \right )\times 10^{-3}$  
      &   $3.3 \times 10^{14}$ \\
$1.3$ &  $\left(0.26^{+0.95}_{-0.93} \right )\times 10^{-4}$ 
      &  $1.4 \times 10^{3}$ \\
$1.5$ &  $\left(-0.10 \pm 0.54 \right )\times 10^{-4}$  
      &  $9.9$ \\
$1.7$ &  $\left(-0.19^{+0.14}_{-0.16} \right )\times 10^{-4}$  
      &  $2.8$ \\
$1.9$ &  $\left(-0.14^{+0.10}_{-0.11} \right )\times 10^{-5}$  
      &  $2.0$ \\
\hline
\hline
$2.1$ &  $\left(0.36^{+0.33}_{-0.30} \right )\times 10^{-6}$  
      &  $0.53$ \\
$2.5$ &  $\left(-0.17^{+0.46}_{-0.15} \right )\times 10^{-6}$ 
      &  $0.14$ \\
$2.9$ &  $\left(-0.22^{+0.33}_{-1.07} \right )\times 10^{-8}$  
      &  $0.16$ \\
\hline
\hline
$3.1$ &  $\left(0.28^{+0.09}_{-0.31} \right )\times 10^{-8}$  
      &  $0.10$ \\
$3.5$ &  $\left(0.58^{+0.18}_{-0.27} \right )\times 10^{-9}$ 
      &  $0.066$ \\
$3.9$ &  $\left(0.85^{+0.58}_{-3.30} \right )\times 10^{-11}$  
      &  $0.072$ \\
\hline
\hline
\end{tabular}
\end{table}

In a previous publication \cite{CKY-long}, we gave the approximate 
limits on $\Lambda_\U$ based on an analysis on 4-fermion contact interactions
\cite{Cheung}.  Those estimates could not take into account
the energy dependence of the unparticle contribution, as shown in 
Eq. (\ref{reduced}).  This is the most important improvement of this work
that we have taken into account the energy dependence of each experimental
set as well as in each event (such as in Drell-Yan production the 
$\hat s$ of each event is different.)

We show in Tables \ref{table1} and \ref{table2} our main results. The 
difference between Table \ref{table1} and \ref{table2} is that the
fittings in Table \ref{table2} are without $q\bar q$ pair production
at LEP2.  This is because we found that when we fitted the unparticle
$\epsilon$ term to the LEP2 $q\bar q$ data alone, the fitted values
showed a $ 3-4\; \sigma$ deviation from zero.  Indeed, it was reported
in Ref. \cite{lep} that the hadronic cross sections 
from $\sqrt{s}=130-207$ GeV are systematically higher (on the average
$1.7\sigma$) than the SM predictions.  Therefore, by combining all the
energy data from LEP2 we obtain the fits $3-4 \;\sigma$ deviation from 
the SM.  
Nevertheless, the combined Tevatron DY, HERA DIS and LEP $q\bar q$ results
do not show any appreciable deviation from the SM, as the LEP2 data are
never dominant.
The lower limits on $\Lambda_\U$ do not change
significantly between Tables \ref{table1} and \ref{table2}.  We also
noticed that at small $d_\U \approx 1.1 - 1.5$ the limit is dominated 
by the HERA data while for larger $d_\U \approx  1.7 - 1.9$ the limit is 
dominated by the Tevatron data.  The 95\% C.L. lower limits on
 $\Lambda_\U$, assuming $\lambda_1 = 1$ 
and $d_\U = 1.1 - 1.9$, are
\begin{equation}
\Lambda_\U = \left \{ \begin{array}{ll}
           1.7 - 5.5 \times 10^{14} \; {\rm TeV} \qquad& \mbox{w/ LEP2 data} \\
           2.0 - 3.3 \times 10^{14} \; {\rm TeV} \qquad & \mbox{w/o LEP2 data} 
                      \end{array}  \right.
\end{equation}
On the other hand, for $d_\U = 2.1 -2.9$ the limits on $\Lambda_\U$ are
\begin{equation}
\Lambda_\U = \left \{ \begin{array}{ll}
          0.14 - 0.62 \; {\rm TeV} \qquad& \mbox{w/ LEP2 data} \\
          0.14 - 0.53 \; {\rm TeV} \qquad & \mbox{w/o LEP2 data} 
                      \end{array}  \right.
\end{equation}
whereas  for $d_\U = 3.1 -3.9$ the limits on $\Lambda_\U$ are
\begin{equation}
\Lambda_\U = 0.066 - 0.10 \;{\rm TeV}  \;.
\end{equation}

If we ignore the requirement of conformality, $d_\U = 1-2$ gives the
most severe bounds on $\Lambda_\U \sim 10^{14} - 2$ TeV, making 
observation at the LHC very difficult.  On the other hand, for 
$d_\U = 2-4$ the bounds are at the electroweak scale ($0.1-1$ TeV), 
making potential observations at the LHC. If conformality is to be
maintained for vector unparticle, the unitarity constraint requires
$d_\U > 3$, for which the bounds from the experimental data 
are very mild, only $0.07 - 0.1$ TeV.  Following Ref.~\cite{strassler}
a consistency condition can be imposed to maintain the conformality
and unitarity
\begin{equation}
  8 \frac{ \sin(-\pi d_\U) \Gamma(2-d_\U) }{(4\pi)^{(2 d_\U -2)} \Gamma(d_\U) }
\frac{2}{\pi} \left( \frac{q^2}{\Lambda_\U^2} \right )^{d_\U} < 1 \;,
\end{equation}
where $q^2 \le \hat s_{\rm max}$ and $\hat s_{\rm max}$ depends on the
experimental conditions. It was shown that \cite{strassler} the bounds
on $\Lambda_\U$ are better constrained by this theoretical argument
than the experimental data for $d_\U \agt 1.5$.  One can extend the
above theoretical condition to $d_\U > 2$, and we obtain, by putting
$\hat s_{\rm max} \simeq (2\; {\rm TeV})^2$ for the Tevatron, 
\begin{eqnarray}
\Lambda_\U & \agt & 1.2 - 2 \;{\rm TeV} \qquad {\rm for} \quad d_\U = 1.1-1.9 
\;,  \nonumber \\
&\agt & 0.6 - 1 \;{\rm TeV} \qquad {\rm for} \quad d_\U = 2.1-2.9 \;,
 \nonumber \\
&\agt & 0.3 - 0.5 \;{\rm TeV} \qquad {\rm for} \quad d_\U = 3.1-3.9 \;.
 \nonumber 
\end{eqnarray}
We can see that bounds from experimental data are more severe for small
$d_\U = 1 - 2$ while the conformality places stronger limits on $\Lambda_\U$
for $d_\U \agt 2$; especially for $d_\U > 3$ the limits are an order of
magnitude stronger than the experimental bounds.

\section{Conclusions}

We have investigated the global constraint on the parity-conserving
$\ell\ell qq $ spin-1 unparticle interactions by comparing the
theoretical predictions given by unparticle against high energy data
sets, including the HERA high-$Q^2$ neutral-current data, Drell-Yan
production at the Tevatron, and the LEPII hadronic cross sections.
Overall, the combined data sets do not show any preference of
unparticle over the SM, although the LEPII data alone does show some
preference.  Thus, we obtain 95\% C.L. lower limits on the unparticle
scale $\Lambda_\U$.  It ranges widely from 2 TeV to $O(10^{14})$ TeV
for $d_\U = 1.1 - 1.9$, whereas from 
$0.14 - 0.62\; (0.066-0.1)$ TeV for $d_\U = 2.1 - 2.9 \; (3.1-3.9)$,
depending sensitively on the unparticle scaling dimension $d_\U$.  
This work, 
by analyzing all the three data sets from scratch, is a
real improvement over the previous naive estimate based on rescaling
the results from the conventional 4-fermion contact interactions.  
The energy dependence of the unparticle contributions in the scattering 
amplitudes are taken into account appropriately.
The unitarity constraint of $d_\U > 3$ for vector unparticle has 
pushed the limit of the unparticle scale $\Lambda_\U$ down to the 
electroweak scale.
Nevertheless, following Ref.~\cite{strassler} we can use the theoretical
condition for conformality and unitarity, and place a useful constraint
on $\Lambda_\U$ for $d_\U > 3$.
In other words, the conformality helps improving the bounds, especially
for $d_\U > 3$.

The limits obtained in this paper serve as the most precise
ones for the parity-conserving spin-1 unparticle interactions.
If parity-violating interactions are included, the parity-violating 
data sets, such as atomic parity violation, have to be taken into
account.  Extensions to the spin-2 unparticle interactions are
straight-forward, but in general the limits so obtained are less
stringent than those from the spin-1 case obtained in this work.

\section*{Acknowledgments}
This research was supported in parts by the NSC
under Grant Numbers 96-2628-M-007-002-MY3 and 98-2112-M-001-014-MY3, 
by the NCTS, by the Boost Program of NTHU and by
the WCU program through the NRF funded by the MEST (R31-2008-000-10057-0).


\begin{thebibliography}{999}

\bibitem{unparticle}
H.~Georgi,
Phys. Rev. Lett. {\bf 98}, 221601 (2007).

\bibitem{Banks-Zaks}
T.~Banks and A.~Zaks,
Nucl. Phys. B {\bf 196}, 189 (1982).

\bibitem{Chen-He}        
S.-L. Chen and X.-G. He,
Phys. Rev. D {\bf 76}, 091702 (2007).

\bibitem{tc}
K.~Cheung, W.~Y.~Keung and T.-C.~Yuan,
  AIP Conf.\ Proc.\  {\bf 1078}, 156 (2009)
  [arXiv:0809.0995 [hep-ph]].

\bibitem{CKY-long}
K.~Cheung, W.~Y.~Keung and T.-C.~Yuan,
Phys. Rev. D {\bf 73}, 075015 (2007). 

\bibitem{bander}
M.~Bander, J.~L.~Feng, A.~Rajaraman and Y.~Shirman,
  Phys.\ Rev.\  D {\bf 76}, 115002 (2007)
  [arXiv:0706.2677 [hep-ph]].

\bibitem{Cheung}
  K.~m.~Cheung,
  Phys.\ Lett.\  B {\bf 517}, 167 (2001)
  [arXiv:hep-ph/0106251].

\bibitem{georgi2}
H.~Georgi,
Phys. Lett. B {\bf 650}, 275 (2007).

\bibitem{CKY-short}
K.~Cheung, W.~Y.~Keung and T.-C.~Yuan,
Phys. Rev. Lett. {\bf 99}, 051803 (2007).

\bibitem{GIR}
B. Grinstein, K. Intriligator and I. Z. Rothstein, 
Phys. Lett. B {\bf 662}, 367 (2008).

\bibitem{lee}
J.-P. Lee, 
arXiv:1009.1730 [hep-ph].

\bibitem{cardy}
V.~Riva and J.~L.~Cardy,
  Phys.\ Lett.\  B {\bf 622}, 339 (2005)
  [arXiv:hep-th/0504197].

\bibitem{mack}
G. Mack, Comm. Math. Phys. {\bf 55}, 1 (1977).

\bibitem{sannino}
F.~Sannino and R.~Zwicky,
  Phys.\ Rev.\  D {\bf 79}, 015016 (2009)
  [arXiv:0810.2686 [hep-ph]].

\bibitem{hera}
S.~Chekanov {\it et al.}  [ZEUS Collaboration],
  Eur.\ Phys.\ J.\  C {\bf 62}, 625 (2009)
  [arXiv:0901.2385 [hep-ex]];
C.~Adloff {\it et al.}  [H1 Collaboration],
  Eur.\ Phys.\ J.\  C {\bf 30}, 1 (2003)
  [arXiv:hep-ex/0304003].

\bibitem{tevatron}
T.~Aaltonen {\it et al.}  [CDF Collaboration],
  Phys.\ Rev.\ Lett.\  {\bf 102}, 031801 (2009)
  [arXiv:0810.2059 [hep-ex]];
C.~Magass  [D0 - Run II Collaboration],
  arXiv:0710.2202 [hep-ex].

\bibitem{lep}
J. Alcaraz {\it et al}, ALEPH Collaboration, DELPHI Collaboration, L3 Collaboration, 
OPAL Collaboration and LEP Electroweak Working Group, 
   arXiv:hep-ex/0612034.

\bibitem{strassler}
A.~Delgado and M.~J.~Strassler,
  Phys.\ Rev.\  D {\bf 81}, 056003 (2010)
  [arXiv:0912.2348 [hep-ph]];

F.~Caracciolo and V.~S.~Rychkov,
  Phys.\ Rev.\  D {\bf 81}, 085037 (2010)
  [arXiv:0912.2726 [hep-th]].
\end{thebibliography}
\end{document}